\documentstyle[procps]{kapproc} %For PostScript fonts
\begin{document}

\articletitle{What fraction of stars formed in infrared galaxies
at high redshift?}

\author{Neil Trentham}
\affil{Institute of Astronomy, University of Cambridge\\
Cambridge CB3 0HA, United Kingdom}
\email{trentham@ast.cam.ac.uk}

\begin{abstract}
Star formation happens in two types of environment:
ultraviolet-bright starbursts (like 30 Doradus and HII
galaxies at low redshift 
and Lyman-break galaxies at high redshift) and  
infrared-bright dust-enshrouded regions (which may be moderately
star-forming
like Orion in the Galaxy or extreme like the core of Arp 220).
In this work I will estimate how many of the stars in the local
Universe formed in each type of environment, using observations
of star-forming galaxies at all redshifts
at different wavelengths
and of the evolution of the field galaxy population.
\end{abstract}


\section{Introduction}

It is now possible to estimate
the star-formation history of the Universe.
This is performed most directly by summing the contributions from
star-forming field galaxies in optical (corresponding to rest-frame
ultraviolet at high redshift) surveys. 
The direct contribution from ultraviolet-bright star-forming
galaxies to the
comoving star-formation rate density is about
$3 \times 10^{-3} \, {\rm M}_{\odot}\,{\rm yr}^{-1}$ ${\rm Mpc}^{-3}$
at redshift $z=0$, rising to
$4 \times 10^{-2} \, {\rm M}_{\odot}\,{\rm yr}^{-1} \, {\rm Mpc}^{-3}$
at $z=1$, before slowly declining to 
$1.5 \times 10^{-2} \, {\rm M}_{\odot}\,{\rm yr}^{-1} \, {\rm Mpc}^{-3}$
at $z=6$ ($h=0.7$, $\Omega_{\Lambda}=0.7$, 
$\Omega_{\rm m}=0.3$; Salpeter IMF; ref.~1).  
This is normally presented in the
uncorrected form of the ``Madau'' or ``Madau-Lilly'' plot.

But galaxies that are forming stars also experience significant dust
extinction -- we know this because 
we see that local star-forming regions like Orion are dusty and
because local spiral galaxies have spectral energy distributions
(SEDs) which peak in the far-infrared.
Correcting for this, Giavalisco et al.~[1] find that the
the total contribution from  
optically selected galaxies to the comoving star-formation rate
density is about
$1.3 \times 10^{-2} \, {\rm M}_{\odot}\,{\rm yr}^{-1} \, {\rm Mpc}^{-3}$
at  $z=0$, rising to
$0.13 \, {\rm M}_{\odot}\,{\rm yr}^{-1} \, {\rm Mpc}^{-3}$
at $z=1$, where it stays roughly constant out to at least $z=6$.
The corrections used come from the analysis of Adelberger \&
Steidel (ref.~2), which is based on multi-wavelength studies of 
a large sample of star-forming
galaxies. 

An additional contribution may comes from extremely dusty galaxies
where the dust is optically thick -- these may be missing altogether
from optical surveys.  Local ultraluminous infrared
galaxies (ULIGs) are examples of this
kind of galaxy; the $V$ extinction
to the core of Arp 220 is $>10$ mag [3]. 
Another example is the host of GRB 010222 [4], which is an
optical sub-$L^*$ galaxy but has a submillimetre star-formation rate of
$\sim 600$ M$_{\odot}$ yr$^{-1}$.  This kind of galaxy may be similar to
the SCBUA galaxies [5] seen in submillimetre surveys
which might have a redshift distribution quite
different from galaxies in optically selected samples.

In this work I assess the contributions from all three of these 
modes of star formation in
generating the current cosmological 
density in stars $\Omega_*$. 
I provide estimates given current observations and outline how
future observations may provide stronger constraints.

\section{Definitions}

The total density in stars in critical units is
$\Omega_* = \Omega_*^{\rm UV} + \Omega_*^{\rm IR/Opt} + \Omega_*^{\rm IR}$

The density of stars seen forming directly in optical galaxies $\Omega_*^{\rm UV}$ 
equals the integral of the uncorrected ultraviolet star formation rate density 
over redshift.
The density of stars that formed in dusty regions within those 
same galaxies $\Omega_*^{\rm IR/Opt}$
equals he integral of the ultraviolet star formation rate density
multiplied by an extinction correction (which may depend on $z$) over redshift.
Finally, the density of stars $\Omega_*^{\rm IR}$
that formed in highly obscured regions whose presence
cannot be inferred from optical observations equals to the infrared star
formation rate density (as measured by SCUBA or in the future ALMA)  
minus the contributions to the previous integral, 
integrated over redshift.

The partition between $\Omega_*^{\rm IR/Opt}$ and $\Omega_*^{\rm IR}$ is somewhat
arbitrary.
Here we put in $\Omega_*^{\rm IR}$ 
all star formation in galaxies whose bolometric luminosity
is greater than some threshold luminosity corresponding to that at which the
optical luminosity no longer tracks the bolometric (mainly far-infrared) luminosity;
locally this happens at a 60-$\mu$m luminosity of 6.3 $\times 10^{10} 
\, {\rm L}_{\odot}$ (ref.~6).
The extinction in these luminous galaxies surely comes from optically thick dust.
Optically thin extinction would tend to happen in galaxies
contributing to $\Omega_*^{\rm IR/Opt}$.

Star formation that happened in ULIGs would be included in $\Omega_*^{\rm IR}$.

The fraction of star formation that is obscured in any particular galaxy  
varies from 0\% to about 99\% [2].
Obscured star formation at the low end of this range mostly
ontributes to $\Omega_*^{\rm IR/Opt}$ 
and at the high end to $\Omega_*^{\rm IR}$.
The host of GRB 010222 would be at the very high end of
this range. 
 
\section{The current cosmological stellar content: what 
needs to be produced}

The time integral of the cosmic star formation rate must equal the luminosity
integral of the galaxy luminosity function:
$$\Sigma_i \int_L \, L \, \phi_i(L) \, \Gamma_i \, {\rm d}L = 
\int_{t(z)} \dot{\rho_*} \, {\rm d}t,
$$
where $\Gamma_i$ is the mass-to-light ratio of stellar population $i$ (this is derived from 
stellar population synthesis models).
From the combination of the SDSS survey measurements at the bright end [7] and CCD mosaic
surveys [e.~g.~ref.~8] at the faint end, the galaxy luminosity function appears to be 
well-described   
by a Schechter function with $M_R^* = -22.0$ and
$\alpha^*=-1.28$ brightward of $M_R=-19$ and a power law with
$\alpha=-1.24$ faintward of $M_R=-19$.
Performing the sum of integrals on the LHS of this equation, 
$$\Omega_* = 0.0036 \pm 0.0020$$ 
in units of the critical density.
About 3/4 of this is in spheroids and 1/4 in disks.
If a Salpeter, not KTG [9] IMF is used,  
$\Omega_*$
is a factor of two higher.

\section{Observational Constraints}

\subsection{Field Galaxy Evolution}

Multi-colour photometry of a near-infrared selected sample of galaxies has permitted
Dickinson, Papovich and colleagues [10] to measure the evolution of $\Omega_*$ with redshift.
Between $z=0$ and $z=1$ they found that about 40\% of the present-day stars in the Universe
formed.  Between $z=1$ and $z=2$, their best-fitting models suggest that a further fraction
$>50$\% formed and only a few percent of the stars that we currently see had formed by
$z=2$.  However, their is considerable uncertainty in the star-formation histories
used in modelling the SEDs of the sample galaxies and the fraction
of stars in place by $z=2$ may be as high as 25\%.  An integral of the extinction-corrected
Madau Plot [1] over redshift from $z=\infty$ up to $z=2$ gives a fraction of about 25\%.
 
Near-infrared surveys [11] have shown that about 30\% of the massive early-type galaxies seen today
were already in place by $z=2$ -- these are the distant red galaxies (DRGs).  Therefore most of the
stars that we believe formed at $z>2$ are not only in massive galaxies today {\it but were
already in massive galaxies by} $z=2$.  If these formed within the large galaxies, they
would have had to form in very extreme bursts. 

\subsection{GRB host galaxies}

Long-duration gamma-ray bursts (GRBs) 
are thought to be linked to the deaths of
massive stars, as suggested by the 
coincident between SN 2003dh and GRB 030329 [12].  Since massive stars do not live long,
this opens up the possibility of using GRB host galaxies as a SFR-selected
sample of galaxies.  

There are, however, complications.
In the context of a collapsar [13] model, GRBs originate preferentially
from stars of high mass and low metallicity.  This means that there will be
more GRBs per unit SFR in galaxies which have a high-mass biassed IMF
or a low metallicity.  Perhaps these two effects
explain the preponderance of GRBs in ULIGs at $z \sim 1$ [14] and
Ly$\alpha$-emitters [ref.~15; see also ref.~16
about the very
low metallicity and high gas column 
density of the host of GRB030323], 
neither of which significantly contribute to $\Omega_*$.

\subsection{Infrared and submillimetre backgrounds and counts}

The extragalactic background light (EBL) is high at infrared wavelength:
$COBE$/DIRBE measured it as $32 \pm 13$  nW m$^{-2}$ sr$^{-1}$ at 140 $\mu$m
and $17 \pm 4$ nW m$^{-2}$ sr$^{-1}$ at 240 $\mu$m [17]. 
The submillimetre background is somewhat lower:
$0.55 \pm 0.15$ nW m$^{-2}$ sr$^{-1}$ at 850 $\mu$m [18].
The optical extragalactic background light also appears to be
high: 12/15/18 nW m$^{-2}$ sr$^{-1}$ at 300/550/800 nm,
with an uncertainty of 50\% [19].  

Madau \& Pozzetti [20] estimated the total EBL as
$55 \pm 20$ nW m$^{-2}$ sr$^{-1}$.  This is lower than the
value of $100 \pm 20$ nW m$^{-2}$ sr$^{-1}$ quoted by Bernstein
et al.~[21], the main difference being an additional component
from the optical background that was previously undetected (however see
ref.~22).

Most of this background is generated by stars, not AGN, else the local
density of supermassive black holes would be overproduced [20].   
Assuming a recycling fraction of 0.4 (much of the material in stars is
returned to the ISM via winds and supernovae; ref.~23), the EBL
implies a stellar density of $\Omega_* = 0.003 - 0.006$, consistent
with the number in Section 3.

One reason the range here is quite large is that the redshift distribution
of infrared star-forming galaxies is  unknowm, and the contribution of
each galaxy to the EBL $\propto (1+z)^{-1}$.
 
The submillimetre background has been resolved and redshifts determined for a
number of bright sources [24].  However, there are indications [25] that the galaxies which
dominate the infrared background are a different population to the galaxies
which dominate the submillimetre background.
The models described by Chary et al.~[25] suggest that these infrared galaxies 
have lower bolometric luminosities and lower redshifts than the ULIGs observed by
Chapman et al.~[24]. 

\subsection{Optical/Infrared observations of star-forming galaxies}

The instantaneous
cosmic SFR at any redshift equals the infrared and optical contributions.
Making extinction corrections to  
star formation rates for high-redshift galaxies is a substitute for direct
measurement at infrared wavelengths.  Only at low redshift $z<1$ can the two be measured
directly.  
The CFRS+$ISO$ survey [26] showed that about (i) 30\% of star formation at $z<1$, 
was visible at optical wavelengths, (ii) most of the remaining 70\% that was in infrared galaxies
was in disturbed systems with red $I-K$ colours, and (iii) about 18\% of the
star formation happened in ULIGs with SFRs in excess of 100 M$_{\odot}$ yr$^{-1}$.
An implication of these results 
is that most of the star formation at $z<1$ happened in infrared galaxies that are neither optical
galaxies with high internal extinction nor ULIGs of the kind seen
by SCUBA.

The first $Spitzer$ results [25] seem to point towards a similar situation at higher redshifts.
The main difference between low and high redshift is that while the redshift distributions of
optical and infrared galaxies are similar {\it within} the $0<z<1$ range, they are very different
at high redshift. 
 
Another implication of these results is that in optically-selected star-forming
galaxies, most of the star formation is being observed directly and these are
at the low end of the obscuration range described by Adelberger and
Steidel [2].
Many of these may be blue compact emission-line galaxies of the type described at this
conference by Lowenthal and Bershady.

\section{The IMF and density in infrared galaxies}

The local star-forming region 30-Doradus has a stellar IMF that 
is Salpeter ($\propto m^{-2.35}$) above 
3 M$_{\odot}$ [27].   
Lyman-break galaxies at high redshift also have
Salpeter IMFs at high masses, an inference based on optical
spectroscopy of cB58 at 
$z=2.7$ [28].
This IMF is attractive in that 
it evolves into the KTG [9] IMF seen locally [29]. 
Unfortunately, no equivalent analysis can be made for infrared
galaxies and it remains a possibility that they have a different
IMF -- if it is high-mass biassed then they will generate
more energy per unit mass of stars formed than will optical
galaxies.

Stars that form in infrared galaxies probably form in
dense star clusters which need to dissipate in order to
produce local galaxies.
Dissipation timescales are long for dense clusters but
are shorter if the cluster is
embedded in a gaseous medium.
Simulations of this physical process [30], along with
inferences about the star-formation history of the Universe from
stellar populations of nearby galaxies [31]
will provide additional constraints on the total amount
of cosmic star formation that occurred in infrared galaxies
and when it happened.

\section{Concluding thoughts}

My current thinking is that
(i) roughly equal amounts of stars formed in each of the following four types
of environments: optically
visible regions, dust-enshrouded regions within optical
galaxies, heavily obscured galaxies with $L_{\rm IR} < 10^{12}$ L$_{\odot}$
and ULIGs with $L_{\rm IR} < 10^{12}$ L$_{\odot}$, and (ii)
most star formation in optically visible regions was in small galaxies 
at all redshifts [32, 33] while most star formation in ULIGs
happened at high redshift, perhaps producing the DRGs.
But these are not strongly held convictions
and I share the optimism felt at this conference that the
puzzle of star formation in galaxies
will be solved over the next few years.

\end{document}